\documentclass[pre,floatfix,a4paper,nofootinbib,amssymb,amsmath,superscriptaddress]{revtex4}
\usepackage{graphicx}
\usepackage{bm}
\usepackage{eucal}
\usepackage{mathrsfs}

\newcommand {\be} {\begin{equation}}
\newcommand {\ee} {\end{equation}}
\newcommand {\Be}{\begin{eqnarray*}}
\newcommand {\Ee} {\end{eqnarray*}}
\newcommand {\bey} {\begin{eqnarray}}
\newcommand {\eey} {\end{eqnarray}}
\newcommand{\bit}{\begin{itemize}}      
\newcommand{\eit}{\end{itemize}}
\newcommand{\bfl}{\begin{flusleft}}
\newcommand{\efl}{\end{flusleft}}
\newcommand{\bfr}{\begin{flushright}}
\newcommand{\bc}{\begin{center}}
\newcommand{\ec}{\end{center}}
\newcommand{\ben}{\begin{enumerate}}    
\newcommand{\een}{\end{enumerate}}

\newcommand{\comment}[1]{}

\newcommand{\kt}{k_\mathrm{B}T}

\newcommand{\xu}{x_\mathrm{u}}
\newcommand{\DG}{\Delta G}

\begin{document} 

\title{Unfolding times for proteins in a force clamp}

\author{Stefano Luccioli}
\affiliation{Istituto dei Sistemi Complessi, CNR, via Madonna del Piano 10, I-50019 Sesto Fiorentino, Italy}
\affiliation{INFN, Sez. Firenze, and CSDC, via Sansone, 1 - I-50019 Sesto Fiorentino, Italy}
\author{Alberto Imparato}
\affiliation{Dept. of Physics and Astronomy, University of Aarhus, 
Ny Munkegade, Building 1520 - DK-8000 Aarhus C, Denmark}
\author{Simon Mitternacht}
\affiliation{Department of Theoretical Physics, Lund University, S\"olvegatan 14A - SE-223 62 Lund, Sweden} 
\affiliation{Department of Informatics, University of Bergen, PB.\ 7803, N-5020 Bergen, Norway }
\author{Anders Irb\"ack}
\affiliation{Department of Theoretical Physics, Lund University, S\"olvegatan 14A - SE-223 62 Lund, Sweden} 
\author{Alessandro Torcini}
\affiliation{Istituto dei Sistemi Complessi, CNR, via Madonna del Piano 10, I-50019 Sesto Fiorentino, Italy}
\affiliation{INFN, Sez. Firenze, and CSDC, via Sansone, 1 - I-50019 Sesto Fiorentino, Italy}

\begin{abstract}
The escape process from the native valley for proteins subjected 
to a constant stretching force is examined using a model for 
a $\beta$-barrel.  For a wide range of forces, the unfolding dynamics 
can be treated as one-dimensional diffusion, parametrized in terms of the 
end-to-end distance.  In particular, the escape times can be evaluated as 
first passage times for a Brownian particle moving on the 
protein free-energy landscape, using the Smoluchowski equation. 
At strong forces, the unfolding process can be viewed as a 
diffusive drift away from the native state, while
at weak forces thermal activation is the relevant mechanism.
An escape-time analysis within this approach reveals a crossover  
from an exponential to an inverse Gaussian escape-time distribution upon 
passing from weak to strong forces. Moreover, a single expression 
valid at weak and strong forces can be devised both for the average unfolding time 
as well as for the corresponding variance. The analysis offers a possible explanation 
of recent experimental findings for ddFLN4 and ubiquitin.
\end{abstract}


\maketitle

Single-molecule pulling experiments have become an important and widely used 
tool for examining mechanical properties of proteins~\cite{forman}. 
These experiments have stimulated a renewed interest in the escape processes
from metastable potential wells in the presence of a 
biasing force~\cite{evans}. Traditionally, the dependence of the 
escape rate $k$ on the stretching force $F$ has often been modeled 
using the phenomenological
Bell formula $k(F)=k_0\,\mathrm{e}^{\beta F\xu}$~\cite{bell}, 
where $\xu$ is the distance from the native to the transition 
state and assumed constant ($\beta=1/\kt$, with $k_\mathrm{B}$ being
Boltzmann's constant and $T$ the temperature). The zero-force rate 
$k_0$ satisfies $k_0\propto\mathrm{e}^{-\beta\DG_0}$, where $\DG_0$ is the
escape free-energy barrier at zero force. There are, however, 
uncertainties about how to extract the zero-force properties 
$k_0$, $\xu$ and $\DG_0$ from observed escape rates at non-zero 
force. One problem is the unknown constant of proportionality in
the expression for $k_0$. Another difficulty is that the 
distance $\xu$ between the native free-energy minimum and the unfolding
barrier, which is assumed constant in the Bell formula, generally depends 
on the applied force.  

To address these problems, several generalizations of the Bell formula 
have recently been proposed~\cite{dudko, liu, friddle, dudko08}. Most of
the extensions are based on the same underlying picture as for the 
Bell formula; the protein is viewed as a Brownian particle moving in a 
tilted one-dimensional potential, $G(x)=G_0(x)-Fx$, where $G_0(x)$ is
the equilibrium free-energy profile. Using different approximations 
and parametrizations of $G_0(x)$, key properties of the escape process 
have been analyzed, like  the mean and variance of the rupture force 
at constant velocity pulling~\cite{dudko,friddle}. It was further  
shown~\cite{dudko08} that the approach of Dudko, Hummer
and Szabo (DHS)~\cite{dudko} is able to describe 
experimentally observed deviations from the Bell formula 
for the protein ddFLN4~\cite{rief}. 

These extensions based on Kramers theory~\cite{review} assume that
the escape barrier is high compared to $\kt$, leading to single-exponential
kinetics. Very recently, Yew et al. analyzed deviations from single-exponential 
kinetics in unfolding simulations based on a C$_\alpha$ model~\cite{paci}.  
By including the next-to-leading term in an eigenfunction expansion, they obtained  
an improved description of the unfolding dynamics at strong force. 
However, a comprehensive picture describing $k(F)$ and the 
full escape-time distribution at both weak and strong force is still missing.     
A key parameter when describing the force dependence is 
the critical force $F_c$, at which the escape barrier disappears.  
In the DHS approach~\cite{dudko,dudko08},   
one has $F_c=\DG_0/\nu\xu$, where $\nu$ is a model parameter ($\nu=1$
corresponds to the Bell formula).  The above-mentioned 
ddFLN4 analysis~\cite{dudko08} (with $\nu=1/2$ or 2/3)  
suggests that $F_c\sim80$--110\,pN for this protein. 
For the titin module I27, on the other hand,   
$F_c$ appears to be significantly larger ($\DG_0/\xu\sim640$\,pN~\cite{dougan}).
Due to different $F_c$, when analyzing experimental data, 
the strong-force regime $F>F_c$ 
may or may not be relevant, depending on the protein. 
 

In this Letter we investigate the response of a model protein to a wide range
of constant pulling forces. We show that, once the free-energy 
landscape is known with sufficient accuracy, 
the usual Smoluchowski equation~\cite{review} 
in one dimension is sufficient to obtain a good estimate of the average
escape time from the native valley and the associated variance.
Two force regimes, separated by the critical force $F_c$, are observed.
For $F<F_c$, unfolding occurs through a thermally activated 
escape process. For $F>F_c$, the unfolding dynamics can instead be 
interpreted as pure diffusion with an external bias. The transition from the
weak- to the strong-force regime is accompanied by a drastic change in the 
shape of the escape-time distribution, from exponential to inverse Gaussian. 
The applicability of this approach to real proteins, at forces studied 
experimentally, is addressed using recently reported data for  
ddFLN4~\cite{rief,dudko08} and ubiquitin \cite{garcia2,garcia1}.

The protein model we consider is the 3D off-lattice
BPN model~\cite{honey,berry,veit}, where each residue is represented
by a single point and is of one of the following three types: 
hydrophobic (B), polar (P) or neutral (N). We study a 46-residue sequence
which is known to form a four-stranded $\beta$-barrel in its native state.
The folding~\cite{honey,berry,veit,guo}
and mechanical unfolding~\cite{cinpull,lit}
of this sequence have been extensively studied.
We analyze via Langevin dynamics the response of this model 
protein to external forces acting on the chain ends in proximity 
of its folding temperature, namely at $T=0.3$~\cite{lit}. 
Parameters values are as in Ref.~\cite{lit}
and all model quantities are dimensionless; 
for a comparison with physical units, see Ref.~\cite{veit}. 

\begin{figure}[h!]
\begin{center}
\includegraphics*[angle=0,width=7.cm]{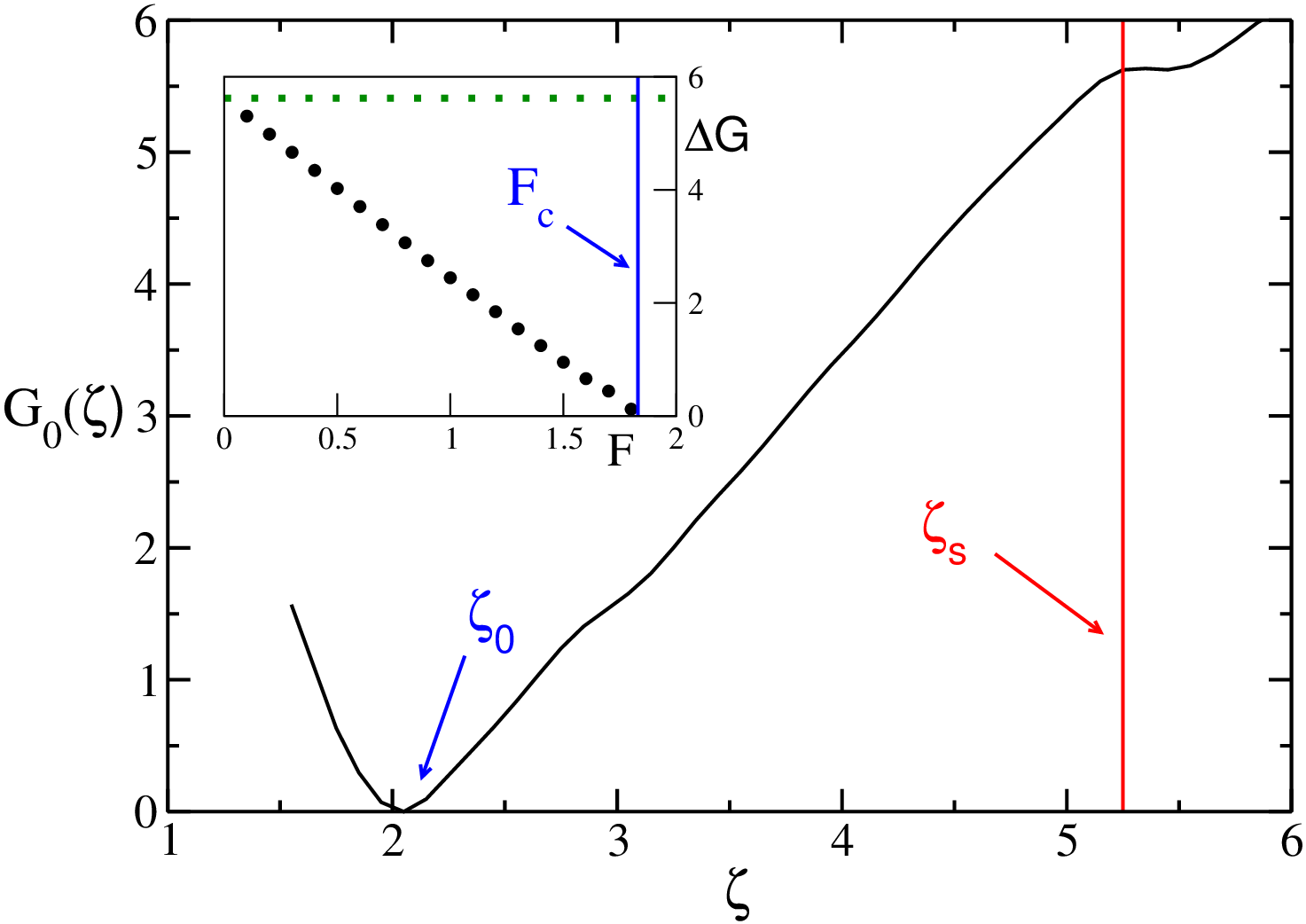}
\end{center}
\caption{(Color online) Free energy $G_0(\zeta)$ at zero force for the 
BPN protein,
calculated as a function of the end-to-end distance $\zeta$. 
The positions of the 
native state, $\zeta_0 \approx 2.0$, and the saddle, $\zeta_s \approx 5.25$,
are indicated. The inset shows the escape barrier 
$\DG$ versus $F$.  The vertical (blue) and horizontal (green) lines
indicate $F_c$ and the zero-force barrier, $\DG_0$.}
\label{free_bar_BPN}
\end{figure}

A typical unfolding trajectory begins with a waiting phase, 
where the end-to-end distance $\zeta$ stays close to its native value.  
This phase is followed by a sudden increase in $\zeta$. A fundamental question is
whether the escape from the native valley can be effectively 
described as one-dimensional diffusion, parametrized in terms of $\zeta$. 
Based on this assumption the unfolding process is commonly
described as the motion of a point-like Brownian particle 
in the potential $G(\zeta)=G_0(\zeta)-F\zeta$, where $G_0(\zeta)$ is the
equilibrium free-energy profile. The average first passage time $\tau(x)$ 
at a threshold $\zeta_s$ for a particle with initial position 
$x \in [\zeta_{0},\zeta_s]$ can be obtained by solving  the 
Smoluchowski equation. One finds that~\cite{review} 
\begin{equation}
\tau(x) = \beta M \gamma \int_{x}^{\zeta_{s}}
\enskip dy \enskip {\rm e}^{\beta G(y)} \int_{\zeta_0}^y \enskip dz \enskip {\rm e}^{- \beta G(z)}
\label{smo}
\end{equation}
where $M$ is the particle mass and $\gamma$ the damping constant. 
The boundaries  at $\zeta_0$ and $\zeta_{s}$ are  
reflecting and absorbing, respectively. When using Eq.~(\ref{smo})
to calculate the escape time from the native valley, $\zeta_0$ is the 
native $\zeta$ and $\zeta_s$ is that of the saddle, 
or barrier, to be crossed. 
The escape time is obtained as $\tau_S\equiv \tau(\zeta_0)$.
In our simulations, escape times are measured 
using a threshold slightly larger than $\zeta_s$,  
to avoid saddle recrossing~\cite{review}. 

We begin by testing the escape-time prediction $\tau_S$ directly 
against simulation results for the BPN protein, 
without making any further assumption on the form of $G(\zeta)$. 
For this purpose, we determine $G(\zeta)$ numerically,  
using methods described in Ref.~\cite{lit}. 
Fig.~\ref{free_bar_BPN} shows the calculated free-energy profile at
zero force, $G_0(\zeta)$, which 
exhibits a pronounced native minimum 
at $\zeta_0\approx2.0$ and a barrier at $\zeta_s\approx5.25$. The height of
the barrier is $\DG_0=G_0(\zeta_s)-G_0(\zeta_0)\approx5.62$. 
The application of a stretching force $F$ 
tilts the free-energy landscape to $G(\zeta)=G_0(\zeta)-F\zeta$ and  
reduces the barrier height $\DG$. As shown in the inset 
of Fig.~\ref{free_bar_BPN}, $\DG$ decreases almost linearly with $F$. 
The barrier finally disappears at $F_c \approx 1.83$. 

\begin{figure}[h!]
\begin{center}
\includegraphics*[angle=0,width=7.cm]{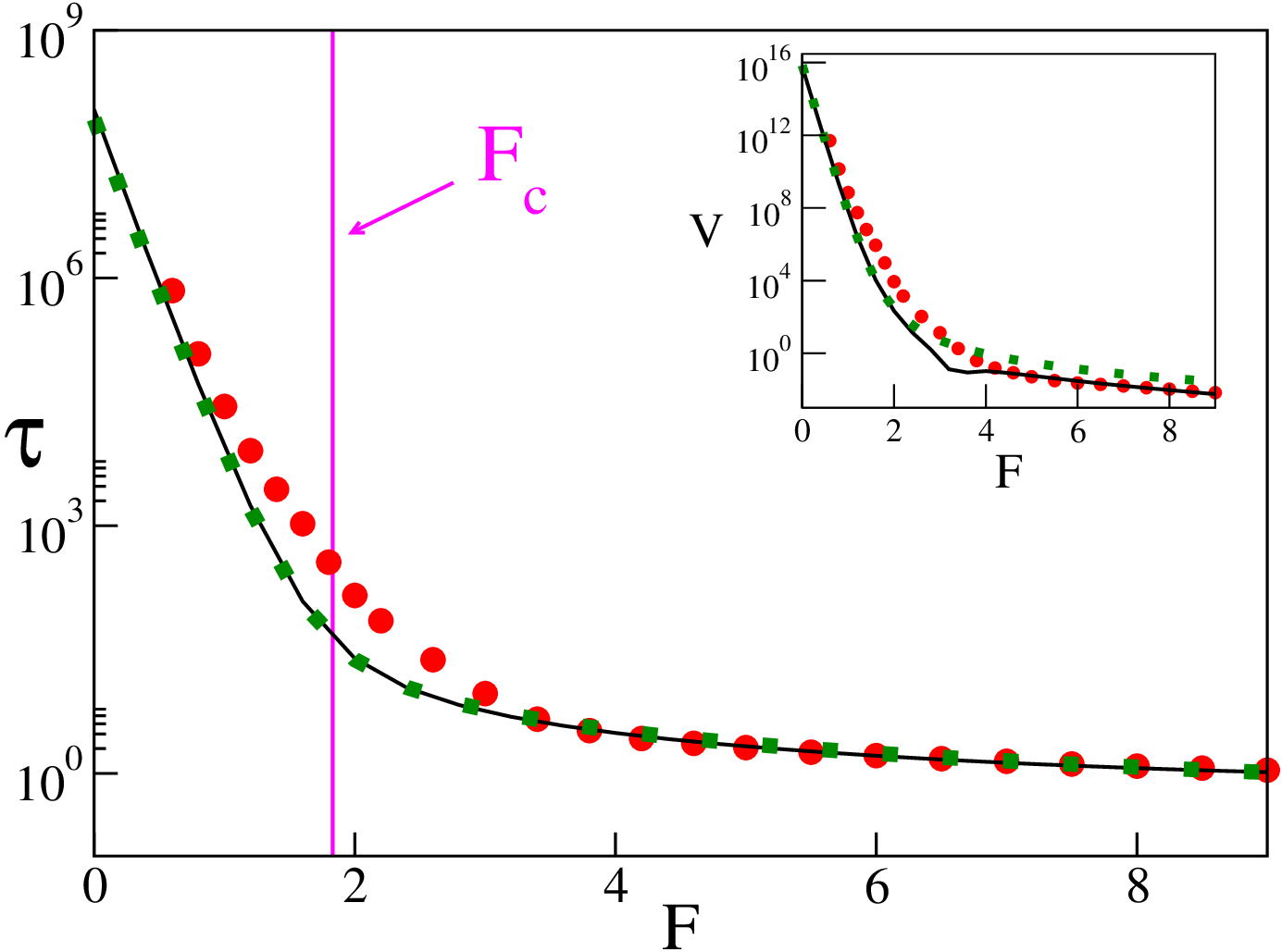}
\end{center}
\caption{(Color online) Average escape time against force 
for the BPN protein. Filled (red) circles are simulation results and the 
(black) curve is the prediction $\tau_S$ obtained from Eq.~(\ref{smo}), 
with $\gamma=0.05$ and $M=46$. 
The vertical (magenta) line indicates $F_c$. 
The dotted (green) line is the estimate $\tau_L$ in 
Eq.~(\ref{redner}), with $M$ and $\gamma$ as above and $a=3.25$. 
The inset shows the variance, $V$.   
Filled (red) circles are simulation results, whereas the (black) curve  
and the dotted (green) line represent the estimates 
$V_S$ and $V_L$, respectively.
}
\label{smol_average}
\end{figure}

Knowing $G(\zeta)$, the escape-time prediction $\tau_S$  
can be obtained by numerically evaluating the double integral in Eq.~(\ref{smo}). 
In Fig.~\ref{smol_average} we compare $\tau_S$ with simulated escape 
times. The agreement is very good for strong 
forces ($F\gtrsim3$) as well as at weak forces ($F \lesssim 1.2$).
Due to computational limitations, it was impossible to investigate forces $<0.6$. 
The regime in which the simulated escape times are most 
difficult to reproduce is around the critical force $F_c$, where there 
is no clear free-energy gradient either towards or away from the native state.   
In this regime, the details of the free-energy profile matter.
It is remarkable, however, that this simple  picture, without employing any 
fitting parameter, is able to describe the behavior at both 
strong and weak forces, despite escape-time differences of almost six orders 
of magnitude.

This analysis, based on the full profile $G(\zeta)$,  addresses in a direct 
manner the question of whether or not the system can be 
described in terms of one-dimensional diffusion. In unfolding 
experiments, $G(\zeta)$ is unknown, and the challenge is to extract the
main features of the free-energy landscape from measured escape times.
This task is greatly facilitated if the free energy can be linearly approximated 
in the interval $[\zeta_0,\zeta_s]$, as
$G(\zeta)=(F_c-F)(\zeta-\zeta_0)$ (up to an additive constant). 
With this approximation, the integrals in Eq.~(\ref{smo}) can be
evaluated analytically. The resulting expression, for the average 
escape time of a diffusive particle in one dimension in the presence 
of a bias ($F$ in the present context), is~\cite{Redner}
\begin{equation}
\tau_L = \frac{M \gamma a}{F-F_c} - \frac{M \gamma k_B T}{(F-F_c)^2} 
\left[1 - {\rm e}^{-\beta (F-F_c) a}  \right]
\label{redner}
\end{equation}
where $a = \zeta_{s} -\zeta_0$ is the distance between the reflecting
and absorbing boundaries.  Unlike the result reported in Ref.~\cite{Redner},  
Eq.~(\ref{redner}) includes the effect of a zero-force barrier, represented by 
the term $F_c a$. The singular terms at $F=F_c$ in Eq.~(\ref{redner}) cancel
out, as they should.  

The assumption that $G(\zeta)$ is linear between the native state and the saddle 
is quite well satisfied for the BPN protein (see Fig.~\ref{free_bar_BPN}).  
Actually, the escape times obtained using this approximation, 
$\tau_L$, essentially coincide with the estimated $\tau_S$ obtained 
using the full $G(\zeta)$, as can be seen from 
Fig.~\ref{smol_average}. Note that Eq.~(\ref{redner}), like 
Eq.~(\ref{smo}), has no parameter that needs to be fitted, because 
we can use the value of $F_c$ previously determined.
While Eq.~(\ref{redner}) well describes the escape time
down to the lowest forces that could be studied, 
one should still be cautious in using this
expression to extrapolate to zero force, because a ``turnover''
to a force-independent process is likely to occur at weak force \cite{Best}.
The extent of this weak-force regime might be non-negligible if
the temperature is high~\cite{Best}.

The variance of the escape time is in the Smoluchowski approach
given by $V_S = \tau_{2,S} - \tau_S^2$, where the second moment 
$\tau_{2,S}$ reads~\cite{gardiner} 
\begin{equation}
\tau_{2,S} = 2 \left(\frac{M \gamma}{k_B T}\right) \int_{\zeta_0}^{\zeta_{s}}
dy\,{\rm e}^{\beta G(y)} \int_{\zeta_0}^y dx\,{\rm e}^{- \beta G(x)} \tau(x)
\label{smo2}
\end{equation}
Like the mean, the variance can be obtained 
analytically if $G(\zeta)$ depends linearly on $\zeta$. This estimate
of the variance, $V_L$, can be found in  Eq.~(S1)~\cite{epapsfile}. 
The inset of  Fig.~\ref{smol_average} shows our simulation results 
for the variance of the escape time for the BPN protein, 
along with the estimates $V_S$ and $V_L$. 
For $ F < F_c$, $V_S$ and $V_L$ are almost identical;  while for $F > F_c$,  
$V_L$ is slightly larger than both $V_S$ and the simulation results, although 
the corresponding three average times are very similar in this regime. 
Overall, both $V_S$ and $V_L$ agree well with the simulation results. 

\begin{figure}[h!]
\begin{center}
\includegraphics*[angle=0,width=7.cm]{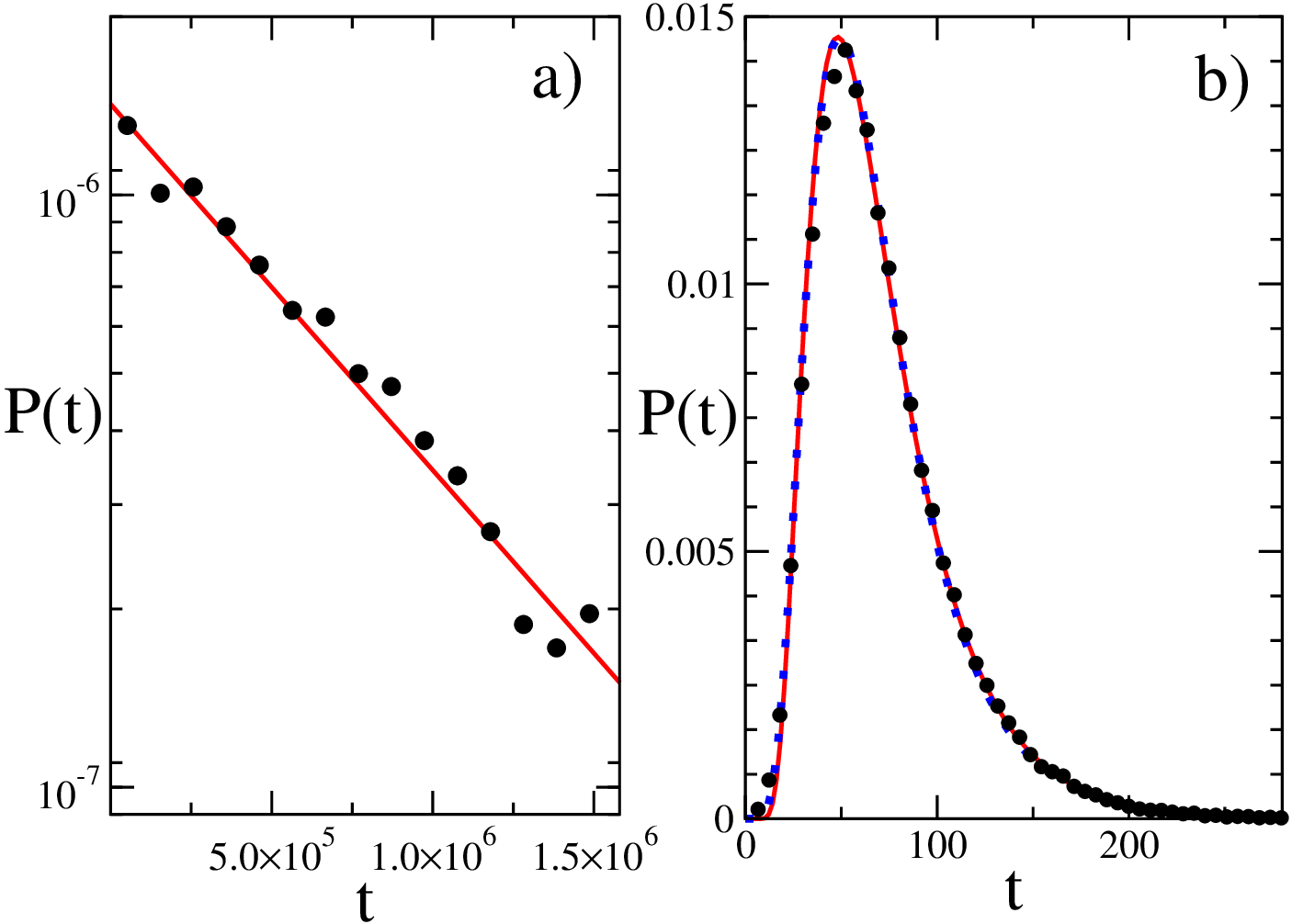}
\end{center}
\caption{(Color online) Escape-time distribution $P(t)$ for the 
BPN protein at two different forces. Large (black) dots are simulation results.
a) $F=0.6\ (<F_c)$.  The (red) curve is an exponential ($\tau = 7.02\cdot10^5$).
b) $F=2.2\ (>F_c)$. The (red) curve is an inverse Gaussian ($\tau= 71.1$, $V=1.35\cdot10^3$),
whereas small (blue) dots represent a log-normal fit to data.
}
\label{distr_BPN}
\end{figure}

It is informative to go beyond the first and second moments and also study 
the full probability distribution of the escape time.  
For $F\lesssim F_c$, we find that the escape-time distribution of the BPN 
protein to a very good approximation is exponential,   
$P(t) = \tau^{-1} \textrm{e}^{-t/\tau}$, with $\tau$ being the mean (see 
Fig.~\ref{distr_BPN}a).  
This observation confirms that at weak forces, where a free-energy barrier is still 
present, the main escape mechanism is thermal activation. At $F\sim F_c$, the 
escape process changes
in character, from a thermally activated process to a diffusive process driven by an 
external bias (force). In the latter regime,  it is known that  first-passage times 
follow a so-called inverse Gaussian distribution~\cite{gaussiana}.
This distribution is given by 
\begin{equation}
P(t)=\frac{\tau}{\sqrt{2 \pi \Gamma t^3}}
{\rm e}^{-(t-\tau)^2/(2 \Gamma t)}
\label{inv_gau}
\end{equation}
where $\tau$ is the mean and $\Gamma= V/\tau$,
$V$ being the variance. This expression indeed provides a very good 
description of our simulation
results at strong forces, as illustrated in Fig.~\ref{distr_BPN}b.
It should be noticed that this comparison does not involve 
any parameter fitting, because $\tau$ and $V$ are determined directly 
from the simulations.  

Previous studies have used a log-normal distribution, rather than the inverse Gaussian,
to describe the escape-time distribution at strong forces~\cite{Cieplak,zamparo}. 
While the log-normal distribution is similar to 
the inverse Gaussian (see Fig.~\ref{distr_BPN}b), 
there is no theoretical background to justify its use in the present context.
The inverse Gaussian distribution is, by contrast,  known to arise from 
biased Brownian motion~\cite{gaussiana}, which provides a simple physical 
picture of the unfolding dynamics at strong forces.  

Having seen that our approach provides a good description of the unfolding 
dynamics of the BPN protein, we now turn to two real proteins, ddFLN4 and 
ubiquitin. Two results of the above analysis are particularly useful when  
comparing with experimental data. The first is Eq.~(\ref{redner}), which 
provides an approximate closed-form expression for the average 
escape time $\tau(F)$ at both weak and strong forces. The second result 
is that the onset of the non-exponential strong-force behavior of $\tau(F)$ 
is accompanied by a change of shape of the escape-time distribution, 
from exponential to inverse Gaussian. 

\begin{figure}[h]
\begin{center}
\includegraphics*[angle=0,width=7.cm]{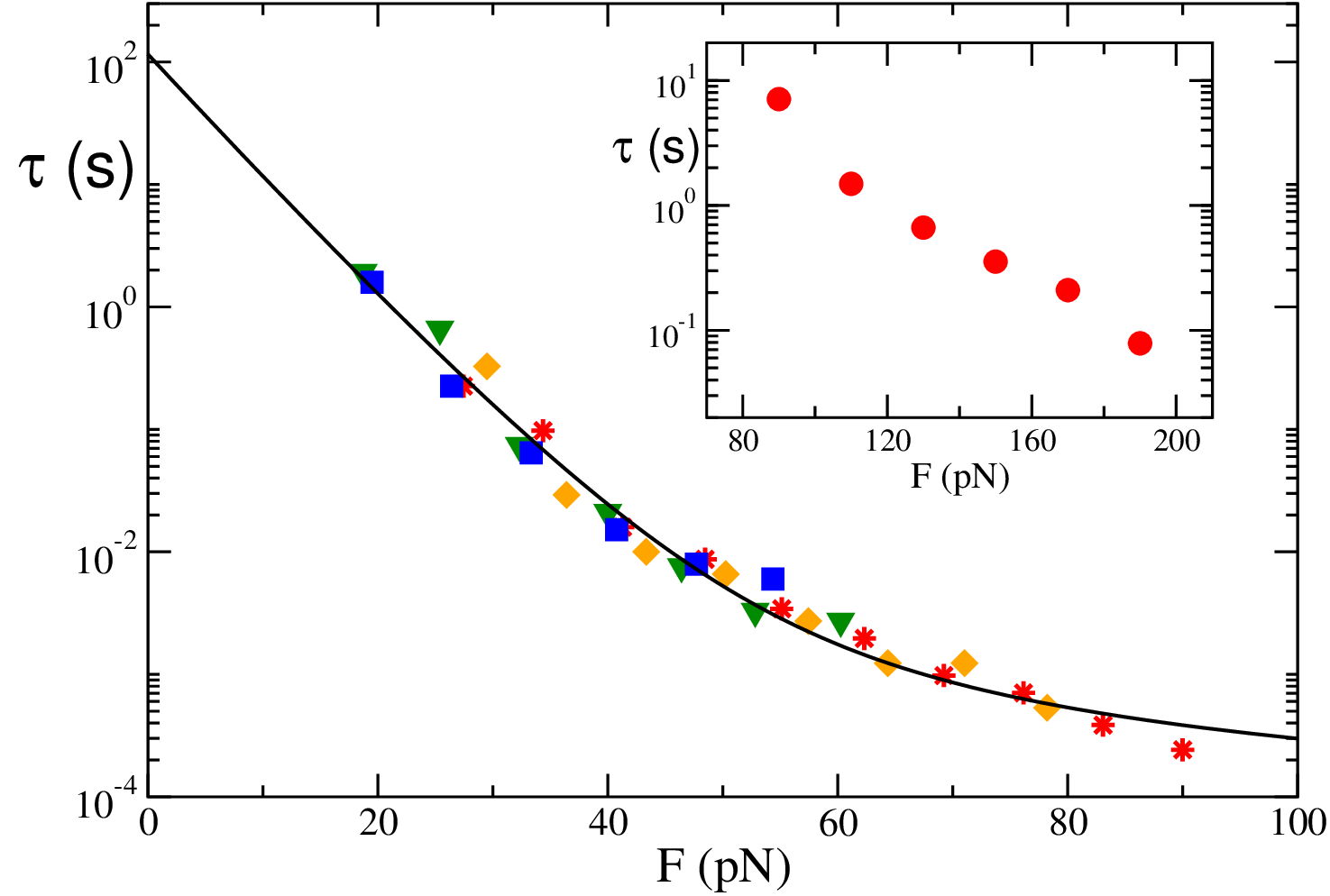}
\end{center}
\caption{(Color online) Average unfolding time versus force for ddFLN4. 
The symbols (explained in detail in Ref.~\cite{epapsfile}) represent experimental 
data, originally obtained 
at constant velocity~\cite{rief} and transformed to constant-force conditions
in Ref.~\cite{dudko08}. The (black) curve is a fit of Eq.~(\ref{redner}) 
($M\gamma=0.012\,\textrm{pN}\cdot\textrm{s}/\textrm{nm}$, $a=1.1$\,nm and 
$F_c = 60$\,pN). The inset shows ubiquitin data from 
AFM force-clamp experiments reported in Ref.~\cite{garcia1}.
}
\label{ubiquitin}
\end{figure}

Experimental unfolding times for ddFLN4 show, as mentioned earlier, 
clear deviations from the Bell formula~\cite{rief,dudko08}. It has been
demonstrated~\cite{dudko08} that the DHS approach~\cite{dudko} 
describes the data well. In Fig.~\ref{ubiquitin},
we show a fit of our Eq.~(\ref{redner}) to the same data. The fit is 
good, and the fitted values $a = 1.1$\,nm (corresponding to
$\xu$) and $\Delta G_0 = F_ca = 9.50$\,kcal/mol are consistent with 
the results of  Ref.~\cite{dudko08}. Unlike the DHS approach, ours 
does not assume the escape barrier to be high. For ddFLN4, 
our fit to the $\tau(F)$ data indicates that the barrier disappears already at  
$F_c\sim60$\,pN. It would be very interesting to see whether the escape-time 
distribution is inverse Gaussian at, say, 100\,pN, but this distribution
has not been evaluated, as far as we know.  

For ubiquitin, the escape-time distribution has been measured
experimentally at 110\,pN~\cite{garcia2}. The data 
were found to be well described by a log-normal 
distribution~\cite{garcia2}, which is very similar to the inverse 
Gaussian one found above at strong forces. Our approach 
thus offers an explanation of the shape of the observed 
distribution. This explanation requires that $F_c<110$\,pN.
Very recent experimental $\tau(F)$ data for 
ubiquitin~\cite{garcia1} show signs of 
deviations from the Bell formula (see inset of Fig.~\ref{ubiquitin}).
However, it was found 
that the data could not discriminate between the Bell and 
DHS formulas~\cite{garcia1}. Neither are the data sufficient 
to permit a stable fit of Eq.~(\ref{redner}), which would 
have given us an independent estimate of $F_c$. The 
assumption that $F_c<110$\,pN seems, however,  fully consistent
with the experimental $\tau(F)$ data. 


In this Letter we have shown for a model protein 
that the unfolding process from the native valley
under force-clamp conditions can be modeled as  
a Brownian motion in a tilted one-dimensional free-energy 
landscape. Moreover, it turned out that this description 
could be further simplified with a surprisingly 
small loss of accuracy, by adopting a linear approximation 
for the free energy. This analysis links deviations 
from the Bell formula for $k(F)$ for
$F>F_c$ to an altered shape of
the escape-time distribution, from exponential to
inverse Gaussian. Comparison with experiments 
indicates that the strong-force regime might set in 
at relatively weak force ($F_c\lesssim100$\,pN) for 
both ddFLN4 and ubiquitin.

\appendix

\newpage

\vskip 0.3 truecm

\centerline{\bf Supplementary Information}

\vskip 0.3 truecm

\centerline{\bf Variance of the unfolding time in the linear approximation}

\vskip 0.3 truecm

\noindent
The variance of the unfolding time as obtained in the Smoluchowski approach is given in 
the Letter, for a general free energy $G(\zeta)=G_0(\zeta)-F\zeta$.  The integrals in this expression 
can be evaluated analytically if the free energy is assumed to be linear in the interval
$[\zeta_0,\zeta_s]$, $G(\zeta)= (F_c-F)(\zeta-\zeta_0)$. At external force $F$, 
the variance is then given by 
\begin{equation}
V_L = \frac{2 a k_B T (M \gamma)^2}{(F-F_c)^3} 
\left(1 + 2 {\rm e}^{-\beta (F-F_c)a}\right)
+ \frac{(M \gamma k_B T)^2}{(F-F_c)^4} 
\left( {\rm e}^{-2 \beta (F-F_c) a} + 4 {\rm e}^{-\beta (F-F_c)a } - 5 \right)
\tag{S1}
\end{equation}
where the symbols are as in the Letter. This result 
extends the one reported in equation (2.2.31) of Ref. \cite{Redner}, for a 
particle diffusing in one dimension in the presence
of a bias ($F$ in our context), to the case
where a barrier (represented by the term $F_c a$) is present also at zero bias.  \vskip 0.3 truecm

\vskip 0.3 truecm

\centerline{\bf Strong force limit}

\vskip 0.3 truecm

\noindent
The expressions for the average unfolding time and the associated variance, 
in the linear approximation, can be simplified for strong forces \cite{tuckwell},
\begin{equation}
\tau_L\rightarrow\tau_{SF}=\frac{M \gamma a}{F-F_c}
\qquad 
V_L\rightarrow V_{SF}=\frac{2 k_B T (M \gamma)^2 a}{(F-F_c)^3}
\tag{S2}
\end{equation}
Using these expressions, it is, in principle, possible to extract 
the critical force ($F_c$) and the distance to the transition state ($a$),
and thereby also the zero-force barrier ($\Delta G_0 = F_c \times a$), 
from escape-time measurements at strong forces. 
 
\vskip 0.3 truecm

\centerline{\bf Description of the experimental data for ddFLN4}

\vskip 0.3 truecm

\noindent
Fig.~4 in the Letter shows unfolding times at different forces for the ddFLN4 protein. The data
have been extracted from Fig.~2b in Ref.~\cite{dudko08}.  The results in
Ref.~\cite{dudko08} were based on constant-velocity AFM pulling experiments     
by Ref.~\cite{rief}. Rupture-force histograms 
obtained by Ref.~\cite{rief} were transformed in Ref.~\cite{dudko08} into 
force-dependent unfolding times 
measurable at constant force. The symbols in Fig.~4 are as in Ref.~\cite{dudko08}
and correspond to different pulling velocities in the original experiments:
$v=200$\,nm/s (blue squares), $v=400$\,nm/s (green triangles);  
$v=2,000$\,nm/s (yellow diamonds), and $v=4,000$\,nm/s (red stars).


\begin{thebibliography}{99}

\bibitem{forman} J.R. Forman and J. Clarke, 
Curr. Opin. Struct. Biol. {\bf 17}, 58 (2007).
 
\bibitem{evans} 
E. Evans and K. Ritchie, Biophys. J. {\bf 72},  1541 (1997).

\bibitem{bell} G. I. Bell, Science {\bf 200}, 618 (1978).

\bibitem{dudko} O.K. Dudko {\it et al.}, \prl {\bf 96}, 108101 (2006).

\bibitem{liu} H.J. Lin {\it et al.}, \prl {\bf 98}, 088304 (2007).

\bibitem{friddle} R. W. Friddle, \prl {\bf 100}, 138302 (2008). 

\bibitem{dudko08} O.K. Dudko {\it et al.},
Proc. Natl. Acad. Sci. U.S.A.  {\bf 105}, 15755 (2008).

\bibitem{rief} M. Schlierf and M. Rief, Biophys. J. {\bf 90}, L33 (2006).

\bibitem{review} P. H\"anggi {\it et al.},
{Rev. Mod. Phys.} {\bf 62}, 251 (1990).

\bibitem{paci} Z.T. Yew {\it et al.}, \prl {\bf 101}, 248104 (2008). 

\bibitem{dougan} L. Dougan {\it et al.}, Proc. Natl. Acad. Sci. U.S.A. {\bf 105}, 3185 (2008).

\bibitem{garcia2} S. Garcia-Manyes {\it et al.}, Biophys. J. {\bf 93}, 2436 (2007).

\bibitem{garcia1} S. Garcia-Manyes {\it et al.}, 
Proc. Natl. Acad. Sci. U.S.A.  {\bf 106}, 10534 (2009).

\bibitem{honey} J.D. Honeycutt and D. Thirumalai,
{Proc. Natl. Acad. Sci. U.S.A.} {\bf 87}, 3526 (1990).

\bibitem{berry} R.S. Berry {\it et al.},
{Proc. Natl. Acad. Sci. U.S.A.} {\bf 94}, 9520 (1997).

\bibitem{veit} T. Veitshans {\it et al.}, {Folding \& Design} {\bf 2}, 1 (1997).

\bibitem{guo} Z. Guo and C.L. Brooks III, {Biopolymers} {\bf 42}, 745 (1997);
J.G. Kim {\it et al.} \prl {\bf 97}, 050601 (2006).

\bibitem{cinpull} F.-Y. Li {\it et al.}, \pre {\bf 63} (2001) 021905;
D.J. Lacks, Biophys. J. {\bf 88} (2005) 3494.

\bibitem{lit} A. Imparato {\it et al.}, \prl {\bf 99}, 168101 (2007);
S. Luccioli {\it et al.}, \pre  {\bf 78}, 031907 (2008). 

\bibitem{Redner}S. Redner, {\it A guide to first passage processes}
(Cambridge University Press, Cambridge, 2001)

\bibitem{Best} R.B. Best {\it et al.},
J. Phys. Chem. {\bf B 112}, 5968 (2008).

\bibitem{gardiner} C.W. Gardiner,
{\it Handbook of Stochastic Methods}
(Springer, Berlin, 2004).

\bibitem{epapsfile} See EPAPS Document No. \# for supplementary material.

\bibitem{gaussiana} R.~S.~Chikara and J.~L.~Folks,
{\it The inverse Gaussian distribution}
(Marcel Dekker, New York, 1988).

\bibitem{zamparo}  A. Imparato {\it et al.},
Phys. Rev. Lett. {\bf 98}, 148102 (2007);
J. Chem. Phys, {\bf 127}, 145105 (2007).

\bibitem{Cieplak} P. Szymczak and M. Cieplak, J. Phys.: Condens. Matter {\bf 18}, L21 (2006)

\end{thebibliography}

\begin{thebibliography}{99}

\bibitem{Redner} S. Redner, {\it A guide to first passage processes}
(Cambridge University Press, Cambridge, 2001).

\bibitem{tuckwell} H.C. Tuckwell, {\it Introduction to theoretical
neurobiology} (Cambridge University Press, Cambridge, 1988)

\bibitem{dudko08} O.K. Dudko {\it et al.},
Proc. Natl. Acad. Sci. U.S.A.  {\bf 105}, 15755 (2008).

\bibitem{rief} M. Schlierf and M. Rief, Biophys. J. {\bf 90}, L33 (2006).

\end{thebibliography}
\end{document}